# Unconventional Magnetism and Band Gap Formation in LiFePO$_4$: Consequence of Polyanion Induced Non-planarity

Ajit Jena[*] and B. R. K. Nanda

*Condensed Matter Theory and Computational Lab*

*Department of Physics, Indian Institute of Technology Madras*

*Chennai, India, 600036*

* e-mail address: ajit@physics.iitm.ac.in

Oxygen plays a critical role in strongly correlated transition metal oxides as crystal field effect is one of the key factors that determine the degree of localization of the valence d/f states. Based on the localization, a set of conventional mechanisms such as Mott-Hubbard, Charge-transfer and Slater were formulated to explain the antiferromagnetic and insulating (AFI) phenomena in many of these correlated systems. From the case study on LiFePO$_4$, through density-functional calculations, we demonstrate that none of these mechanisms are strictly applicable to explain the AFI behavior when the transition metal oxides have polyanions such as (PO$_4$)$^{3-}$. The symmetry-lowering of the metal-oxygen complex, to stabilize the polyanion, creates an asymmetric crystal field for d/f states. In LiFePO$_4$ this field creates completely non-degenerate Fe-d states which, with negligible p-d and d-d covalent interactions, become atomically localized to ensure a gap at the Fermi level. Due to large exchange splitting, high spin state is favored and an antiferromagnetic configuration is stabilized. For the prototype LiFePO$_4$, independent electron approximation is good enough to obtain the AFI ground state. Inclusion of additional correlation measures like Hubbard U simply amplifies the gap and therefore LiFePO$_4$ can be preferably called as weakly coupled Mott insulator.

## INTRODUCTION

LiFePO$_4$ (LFPO) is considered to be one of the most efficient cathode materials.[1,2] It offers reasonably high operating circuit voltage, one-dimensional Li ion diffusion and large capacity.[3,4] Therefore, most of the research activities on LFPO are about its electro-chemical properties. However, electronic and magnetic structure of LFPO is also equally interesting as it exhibits antiferromagnetic and insulating (AFI) behavior[5-8] similar to many of the strongly correlated transition metal oxides (TMO)[9]. Like many TMOs (e.g. NiO, FeO),[9,10] LFPO has a band gap problem. Without appropriate correlation measures, either the band gap is found to be absent or very small[11-14] and with correlation measures the band gap is predicted to be large (~ 3.5 eV)[13,15,16] which matches well with the experimental value.[16] Theoretical explanation of the magnetic properties of LFPO is also equally diverse. It is predicted to be either an antiferromagnetic Mott-insulator[17] or a ferromagnetic metal[11,13] or a ferromagnetic half-metal.[12] Despite of having many virtues of regular TMOs, the structure of LFPO differs significantly. Firstly it involves a stable polyanion (PO$_4$)$^{3-}$ and secondly the Fe-O complex is highly asymmetric as the expected planar and octahedral symmetries are completely broken.

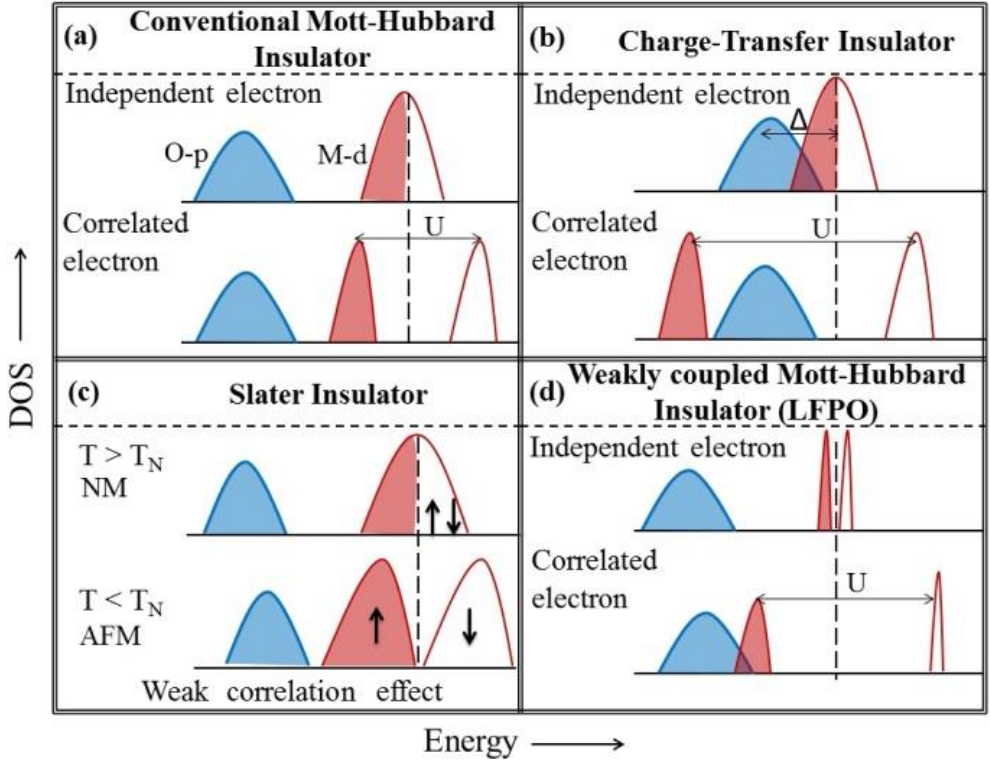

**Figure 1.** Schematic illustration of several mechanisms leading to AFI behavior in TMOs. Strong correlation effect creates Mott-Hubbard[10,18] and Charge transfer type insulators[19]. Slater insulator is driven by Neel temperature.[20] Very weakly coupled d states in bulk LFPO form the gap at the Fermi level as in low dimensional systems such as clusters and nano dots.[21] The band gap is amplified with the inclusion of Hubbard U.

In this paper, we have performed density functional calculations on LFPO to investigate the link between the presence of the polyanion, Fe-O asymmetric complex and the AFI behavior. Based on the results we propose a mechanism which is schematically illustrated in Fig. 1(d). We expect that LFPO is just a prototype and many other TMOs with polyanions may follow this mechanism to obtain their AFI ground state. For the purpose of comparison we have also illustrated the Mott-Hubbard,[10,18] Charge-Transfer[19] and Slater[20] mechanism, respectively in Fig.1a -c, which are appropriate to explain the AFI nature of many strongly correlated TMOs.

The symmetric M-O complexes in TMOs splits the five-fold degenerate atomic d states to new degenerate d-states through crystal field splitting.[22] For example the octahedral M-O complex creates triply degnerate $t_{2g}$ and doubly degenerate $e_g$ states. The new degenerate states undergo coordinated covalent interactions with the O-p states to become localized or itinerant.[9] Due to strong correlation effect, the localized states, depending on occupancy, either lead to Mott insulators (Fig. 1a) ( e.g. LaVO3, MnO, FeO)[9,10] or lead to Charge-transfer insulators (Fig. 1(b)) (e.g. NiO, CuO).[10,19,23]

There are some other TMOs, mostly involving weakly localized d electrons (e.g. $Sr_2IrO_4$),[24] which undergo metal-insulator transition at the Neel temperature ($T_N$). The antiferromagnetic ordering stabilizes as electrons with opposite spins move in different potentials[20] below $T_N$. As a consequence each Brillouin zone is reduced by half and each energy level splits into two with a gap in the middle as shown in Fig. 1c.

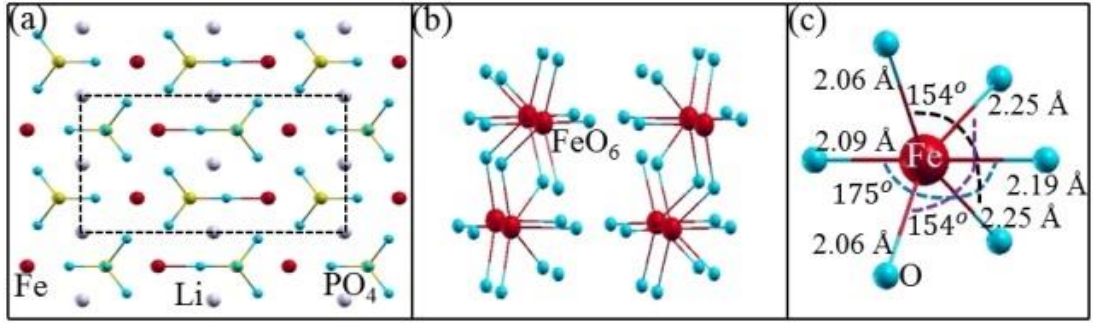

**Figure 2.** (a) Olivine crystal structure of LiFePO$_4$ viewed from the 001 plane. It shows as if Fe and Li ions are distributed in a matrix of PO$_4$ tetrahedra. The conventional unit cell is shown by the rectangle with dashed line. (b) The FeO$_6$ complexes tilted with each other and are also rotated with respect to the crystal axes. (c) Demonstration of octahedral symmetry breaking through unequal Fe-O bond lengths and O-Fe-O bond angles.

| Lattice Const. (Å) | | Fe (4c) | P (4c) | O1 (4c) | O2 (4c) | O3 (8d) |
|---|---|---|---|---|---|---|
| Expt | 10.227 | 0.29 | 0.09 | 0.09 | 0.45 | 0.17 |
|      | 6.004  | 0.25 | 0.25 | 0.25 | 0.25 | 0.05 |
|      | 4.692  | -0.03 | 0.43 | 0.75 | 0.19 | 0.30 |
| DFT  | 10.201 | 0.27 | 0.09 | 0.09 | 0.45 | 0.16 |
|      | 5.990  | 0.25 | 0.25 | 0.25 | 0.25 | 0.04 |
|      | 4.724  | -0.03 | 0.41 | 0.74 | 0.21 | 0.28 |

**Table 1.** Experimental[25] and DFT optimized structural parameters of LFPO. Li occupies 4a (0 0 0). The DFT results are obtained within GGA+U (U = 3eV).

In the case of LFPO, we show that the planar and octahedral symmetries among Fe and O ions are lost as phosphorus strongly attracts the oxygens to form stable (PO$_4$)$^{3-}$ tetrahedral polyanions in the system. This lowering in symmetry introduces a completely anisotropic and inhomogeneous crystal field to create multiple non-degenerate d-states which are devoid of any covalent interaction with the O-p orbitals. These atomic like d-states ensure a gap at the Fermi level to drive the insulating behavior in LFPO as we see in zero dimensional systems such as clusters and nano dots.[21] Also these states undergo large spin-splitting below T$_N$ to create multiple spin half states which mediate Heisenberg type antiferromagnetic interaction in the system. The correlation correction, made through Hubbard U, simply increases the magnitude of the gap as demonstrated in Fig. 1d.

## Structural and Computational Details

*Olivine Crystal Structure*: Crystal structure of LiFePO$_4$ is orthorhombic and it belongs to the Olivine family of compounds[1,25] with space group Pnma (No. 62). The two deterministic features of the crystal structure of LFPO are: (i) the presence of perfect PO$_4$ tetrahedras (Fig. 2a) and (ii) the presence of FeO$_6$ complexes (Fig. 2b), where the octahderal symmetry is lost due to unequal bond lengths and bond angles as shown in Fig. 2c. The crystal axes differ from the axes of the FeO$_6$ complexes and neighboring FeO$_6$ complexes are tilted with each other.

*Computational Details:* Density functional calculations are performed using the Vanderbilt ultra-soft pseudo-potentials[26] and plane wave basis sets as implemented in Quantum Espresso (QE).[27] Exchange-correlation potential is approximated through PBE-GGA functional.[28] Some of the calculations are performed using LDA for comparison and analysis purpose. Strong correlation is a natural phenomena in transition metal oxides. To account for this, parameterized Hubbard U is included in our *ab initio* calculations. The kinetic energy cutoff to fix the number of plane waves is taken as 30 Ry. However, the kinetic energy cutoff for charge density is taken as 250 Ry. A 6x10x12 k-mesh of the BZ, yielding 456 irreducible k-points, for the regular unit cell is found to be sufficient to calculate the total energy with reasonable accuracy within pseudopotential approximation. Table 1 shows that the DFT optimized structure is quite close to the experimental one. Hence it is expected that the DFT calculations will reproduce the low temperature experimental properties of LFPO.

## RESULTS

### Non-Planarity and Structural Stability

Crystals with planar geometry are highly symmetric and therefore many compounds, particularly the TMOs prefer to stabilize in a planar (layered) crystal structure. Some of the well known planar TMOs are monoxides like MnO,[29] NiO,[30] perovskites (AMO$_3$),[31,32] cuprates[33] and Ruddlesden-Popper series: A$_{n+1}$M$_n$O$_{3n+1}$.[34,35] In these TMOs, the transition metal and oxygen ions lie (almost) on a plane. As a consequence, the O ligands of the M-O complex create a symmetric crystal field to split the five-fold degenerate atomic M-d states into multiple degenerate states. For example if there is an octahedral symmetry, as in the case of perovskites, the d-states split into triply degenerate t$_{2g}$ and doubly degenerate e$_g$ states.[22,36] These degenerate states, depending on their occupancies and strong correlation due to localization, exhibit many exotic phenomena. These include half-metallicity,[37] itinerant magnetisms,[38] colossal magneto resistance (CMR),[39] Mott-Hubbard and Charge-transfer insulators[10,19] and high-T$_C$ superconductivity.[40]

Unlike the TMOs discussed above, LFPO is highly non-planar, yet stable and shows AFI behavior. Therefore, it is paramount to study the link between non-planarity and structural stability which, in return, will give useful insight to the electronic and magnetic structure of this compound. In this context, we have carried out a virtual structural deformation experiment as demonstrated in Fig. 3. We start with a perfectly planar hypothetical structure and gradually deform it to the experimental non-planar structure and in each step we performed the *ab initio* calculations. The hypothetical planar structure may be realized experimentally through atomic layer deposition (ALD) and molecular beam epitaxy (MBE) approaches.

Though for the virtual experiment several intermediate structures between the planar and the experimental structure were studied, only three of them, whose Wyckoff positions are listed in Table 2, are discussed here to avoid the redundancy. The stability of these structures are measured through the total energy calculations using the ground state antiferromagnetic ordering of the experimental structure.

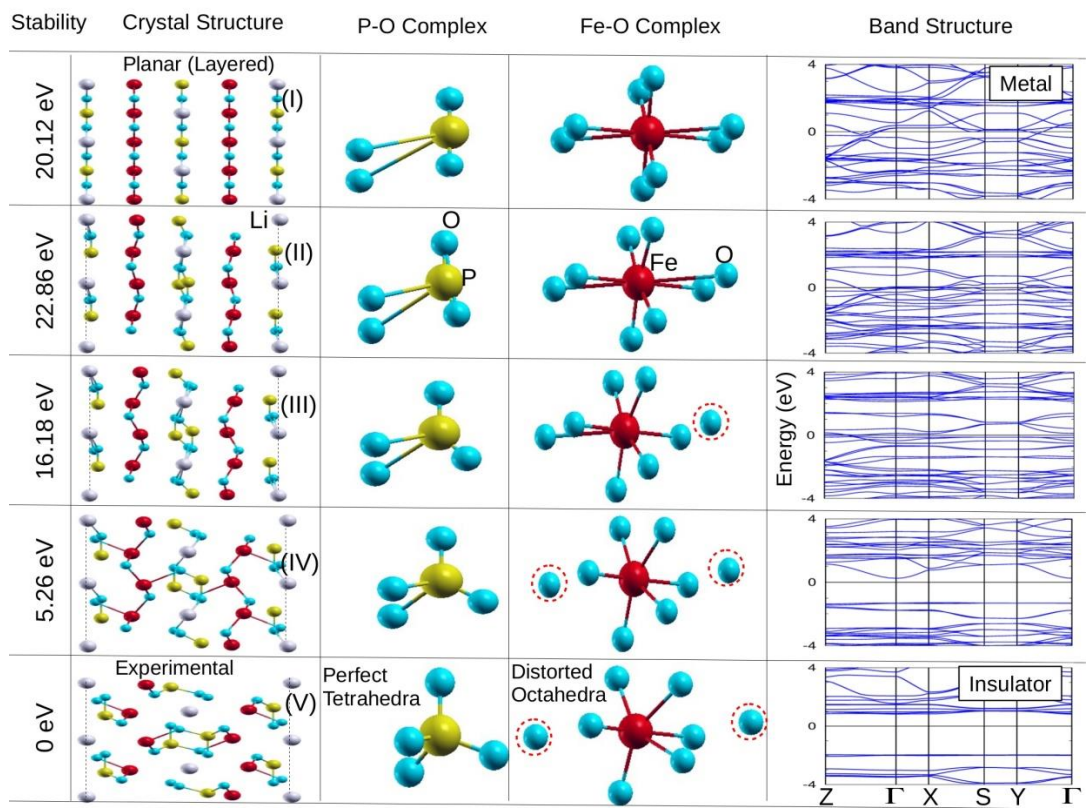

**Figure 3.** Based on the virtual experiment, gradual transition from a perfectly planar (layered) hypothetical structure (top) to the experimental non-planar structure (bottom) of LFPO. The Wyckoff positions for each of the structures are listed in Table 2. The stability, in eV, of each structure is mentioned in the extreme left panel. The planar deformation is shown in the second panel. Formation of perfect $PO_4$ tetrahedra and largely distorted $FeO_6$ complexes concurring to the planar deformation are displayed in third and fourth panels respectively. The corresponding band structures are plotted on the extreme right panel. The red dotted circles indicate the detachment of O ion from the parent complex. The Fermi energy ($E_F$) is set to zero for band structures plotted here and also in remaining relevant figures. In the starting hypothetical $FeO_8$ complex four of the Fe-O (in plane) bonds have length close to 1.9 Å, the other four (inter plane) Fe-O bonds have length close to 2.8 Å.

A comparison of the structures and the corresponding total energies (Fig. 3) suggests that the lowering in symmetry through planar deformation brings stability to the structure. In fact the planar structure is unstable by ~ 20 eV, with respect to the experimental structure, which is too high for a crystal. During the process of deformation, $FeO_8$ complexes give rise to $FeO_6$ complexes and concurrently perfect tetrahedral $(PO_4)^{3-}$ polyanions are formed. The octahedral symmetries, $C_3$, $C_2$, $C_4$ and $C_2'$[22] are far-off for the $FeO_6$ complexes since we have highly unequal Fe-O bond lengths and O-Fe-O bond angles are significantly deviated from the ideal $180^0$.

The relation between structural stability and formation of $PO_4$ tetrahedra is quite evident from the energy comparison of structure III and IV. As we move from III to IV, the $PO_4$ complex takes the shape of a tetrahedra by attracting two O ions from the $FeO_8$ complex. In this process the system gains stability by ~ 11 eV. Further perfection of the tetrahedral shape (structure V) makes the system most stable. The consequence is very significant in realizing LFPO as a cathode material. As Li ion has minimal role for the structural stability, it can be easily diffused and absorbed to facilitate charging and discharging process respectively. The non-planarity brings a big change in the electronic and magnetic properties of LFPO as well. For the planar structure (I), i.e. in the absence of PO$_4$ tetrahedra, the band structure is metallic with widely dispersed bands crossing the Fermi level ($E_F$). By lowering the symmetry the bands at $E_F$ become less and less dispersed. For the experimental structure they are almost flat to induce insulating behavior in this compound. Detailed discussion on the mechanisms that lead to the AFI behavior in LFPO is made in the following two sections. We note that weak electron conductivity is one of the major disadvantages of LFPO as cathode material.[7,16] Bottom two structures in Fig. 3 suggest that a small distortion in tetrahedra enhances the band dispersion substantially. The distortion may be achieved by external pressures or doping.

| Structure | Li (4a) | Fe (4c) | P (4c) | O1 (4c) | O2 (4c) | O3 (8d) |
|---|---|---|---|---|---|---|
| I (Hypothetical, Planar) | 0 | 0.25 | 0 | 0 | 0.5 | 0.25 |
| | 0 | 0.25 | 0.25 | 0.25 | 0.25 | 0 |
| | 0 | 0 | 0.50 | 0.75 | 0.25 | 0.25 |
| II | 0 | 0.25 | 0 | 0 | 0.5 | 0.25 0.05 **0.3** |
| | 0 | 0.25 | 0.25 | 0.25 | 0.25 | |
| | 0 | **-0.03** | **0.43** | **0.75** | **0.19** | |
| III | 0 | **0.26** | **0.02** | **0.0225** | **0.4875** | **0.23** 0.05 0.3 |
| | 0 | 0.25 | 0.25 | 0.25 | 0.25 | |
| | 0 | -0.03 | 0.43 | 0.75 | 0.19 | |
| IV | 0 | **0.27** | **0.04**0 | **0.045** | **0.475** | **0.21** 0.05 0.3 |
| | 0 | 0.25 | .25 | 0.25 | 0.25 | |
| | 0 | -0.03 | 0.43 | 0.75 | 0.19 | |
| V (Experimental) | 0 | **0.29** | **0.09** | **0.09** | **0.45** | **0.17** 0.05 0.3 |
| | 0 | 0.25 | 0.25 | 0.25 | 0.25 | |
| | 0 | -0.03 | 0.43 | 0.75 | 0.19 | |

**Table 2**. The Wyckoff positions for the structures shown in Fig. 3. The numbers written in bold indicate the changes made with respect to the previous structure.

## Electronic and Magnetic Ground State

The objective of this section is to see how octahedral asymmetry due to structural non-planarity affects the electronic and magnetic properties of LFPO. It is well known that a small distortion of the MnO$_6$ octahedra breaks the $e_g$ degeneracy in LaMnO$_3$ and as a consequence A-type AFI ground state emerges in this compound.[36] However, in most of the theoretical studies[11,13,14,15] on LFPO the primary structural assumption is that the FeO$_6$ complex is not distorted to the extent that $t_{2g}$ and $e_g$ symmetries of the Fe-d states are broken. While such an assumption does not affect the energetics and hence stability of the system, it lacks in explaining the electronic and magnetic behavior. Also, as LFPO has 3d electrons, it is necessary to investigate the correlation effect on the electronic properties of this system. To our knowledge there are few literature which have partly discussed the correlation effect using DFT+U calculations[13,15-17] and using DFT+ dynamical mean-field theory (DMFT) calculations.[41,42] While the DMFT studies are restricted to the paramagnetic phase, DFT+U calculations were carried out to match the theoretical band gap with the experimental band gap.[15,16] Therefore, a definitive mechanism to explain the experimental AFI behavior has not evolved so far.

*Electronic structure of LiFePO$_4$ within GGA*

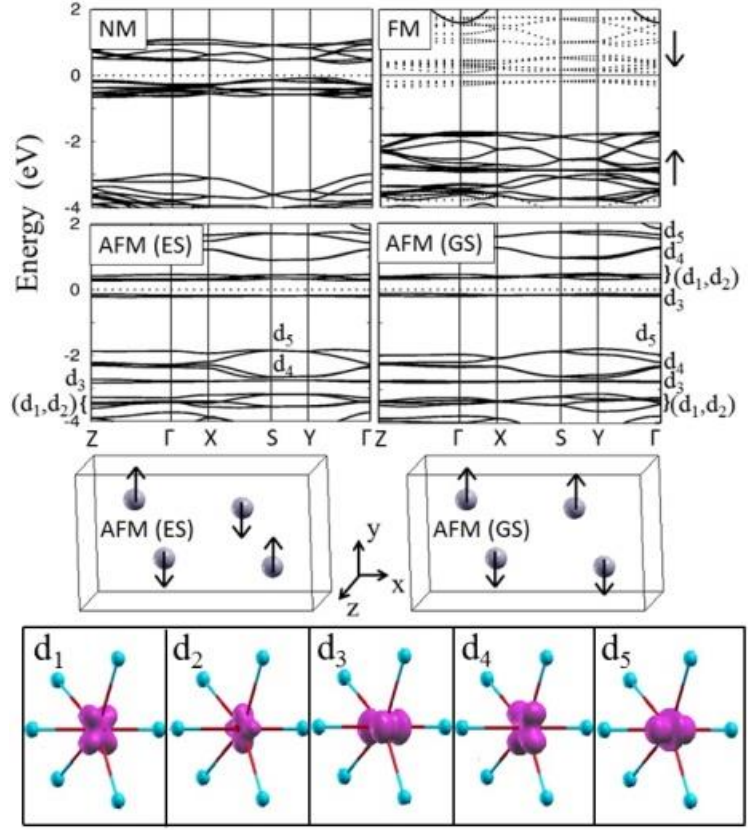

**Figure 4.** Upper: DFT-GGA obtained LFPO band structure of non-magnetic (NM), ferromagnetic (FM) and antiferromagnetic (ground state (GS) and first excited state (ES)) configurations. For the FM case, solid and dashed lines respectively represent the band structure of spin-majority and minority channels. The asymmetric crystal field splits the d-states into multiple non-degenerate states $d_1$, $d_2$, $d_3$, $d_4$, $d_5$. Middle: The ground state and excited state antiferromagnetic spin arrangement. Bottom: The three dimensional electron density map plotted, for one of the FeO$_6$ complex, in different energy ranges to reflect the shape of the non-degenerate d states in AFM (GS) structure. The value of the iso-surface was set to 0.02 e/Å$^3$.

Electronic structure of LFPO can be well understood from Fig. 4 we have plotted the GGA band structures near E$_F$ for the non-magnetic, ferromagnetic and two antiferromagnetic (ground and first excited states) configurations. The band structures reveal three universal phenomena: (a) existence of a narrow band gap at E$_F$, (b) non-dispersive bands in the vicinity of E$_F$ and (c) Fermi level is occupied by the Fe-d states while O-p states lie below the d states (not shown in the figure). These observations, combined together, are in contradiction with any of the conventional magnetic insulators such as Slater, Mott and Charge-transfer which are schematically illustrated in Fig. 1. The non-magnetic insulating feature of the band structure agrees well with the fact that experimentally LFPO is found to be insulating far beyond the Neel temperature.[7,8] Therefore, it contradicts the Slater mechanism.[20] As the band gap exists even without additional correlation (Hubbard U), it is not appropriate to define LFPO as a conventional Mott insulator.[18] The charge-transfer mechanism[19] is ruled out as we have d-d gap instead of a p-d gap at $E_F$.

The band structures in Fig. 4 further reveal that irrespective of the magnetic order, non-dispersive bands at $E_F$ are basically non-degenerate Fe-d states. The distinction of the non-degenerate states is very prominent in the AFM structures. The three dimensional electron density of these non-dispersive bands for the AFM(GS) structure, plotted in the bottom panel of Fig. 4, clearly shows that these bands are basically the individual atomic d orbtials. Therefore, the p-d covalent interaction is either absent or negligible which can be further confirmed from the bandwidth of these states.

A careful observation of Fig. 4, tells us that the band width of the non-degenerate states lies in the range of 0.24 to 0.4 eV for the unstable NM, 0.22 to 0.84 eV for the FM configurations and 0.08 to 0.52 eV for the stable AFM configuration. If we map these bandwidths to that of a oversimplified nearest-neighbor tight-binding band dispersion ($2t \cos ka$) for a one-dimensional lattice of length 'a', then the hopping parameter will lie in the range 0.06 to 0.1eV for NM, 0.055 to 0.21 eV for FM and 0.016 to 0.13 eV for AFM configuration. Such a hopping strength is very negligible to assume any covalent p-d or d-d interactions in the system.

The non-interacting d-orbitals in the presence of a weak crystal field are filled following the Hund's rule and hence $Fe^{2+}$ favors high spin (HS) state. This agrees well with the DFT prediction as the AFM (GS) band structure shows that except four spin-minority d states, the rest are lying below $E_F$. In fact the NM configuration represents the low spin (LS) state for $Fe^{2+}$ and is highly unstable, approximately by 0.5 eV, than that of the magnetic configurations. At higher temperature it will give rise to paramagnetic insulating phase. We note that for odd number of d electrons ($Fe^{3+}$), as in the case of the de-lithiated compound $FePO_4$, the hypothetical NM phase may have states pinned on the Fermi level, but the paramagnetic phase is always insulating. This is discussed in detail in Fig. 1 of the supplementary material attached with this paper. Contrary to the experimental observations of HS configuration (S = 5/2), the DMFT results[41] provide an intermediate-spin state (S = 3/2) for $FePO_4$ in the paramagnetic insulating phase.

*Effects of Exchange Correlation and Onsite Correlation on the Electronic Structure*

Appropriate exchange correlation functionals are vital for accurate DFT prediction of the electronic structure of solids, particularly for the correlated TMOs due to presence of localized d states. A simple case study is the family of monoxides. While LDA predicts accurately the ground state magnetic ordering for NiO and MnO, it gives a smaller band gap than the experimental values.[10,43] For FeO and CoO, LDA suggests a metallic solution while in reality these are insulators with wide band gaps.[43] The errors are attributed to the inability of LDA to account for the correlation. One way to correct the error is through a Hubbard term $H_1$.[10]

$$H_1 = \frac{1}{2} U \sum_{\substack{\nu\nu' \\ (\nu \neq \nu')}} n_{i\nu} n_{i\nu'} \ , \qquad (1)$$

where $i$ represents the site index and $v = (m, \sigma)$ collectively represents the orbital (m) and spin index ($\sigma$) of the state. U serves the purpose of onsite repulsion between the states $v$ and $v'$. Within mean field solution, the pairs $v$ and $v'$ split to produce a gap ~ U between them. While one of the states gets occupied, other remains above $E_F$. Based on the strong correlation effect, two of the mechanisms (Mott-Hubbard and Charge-transfer) that have evolved to explain the AFI behavior in TMOs are schematically shown in Fig.1.

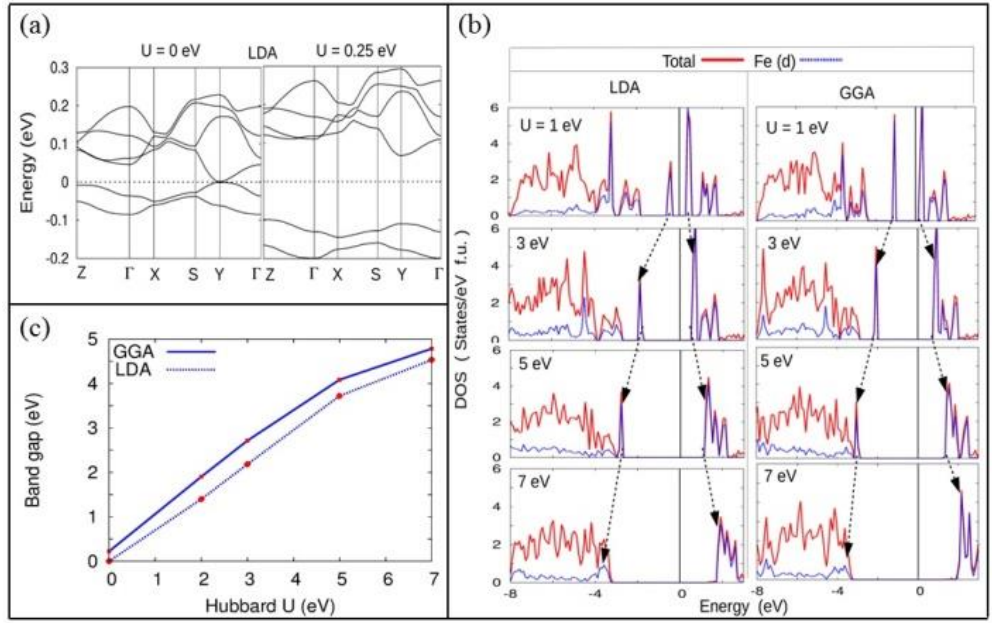

**Figure 5.** (a) Band structure of LFPO at $E_F$ within LDA shows a zero band gap (left). A small value of U is sufficient to open the gap (right). (b) Ground state antiferromagnetic DOS for different values of Hubbard U as obtained using LDA (left) and GGA (right). The arrows indicate the gradual widening of the gap as well as the penetration of Fe-d states in the O-p spectrum as U increases. Band gap with respect to U is quantified in (c).

Earlier DFT calculations[11-13] using LDA predict ferromagnetic and metallic ground state for LFPO. Inclusion of U makes it AFI and the gap resembles to that of a Mott insulator.[17] However, the electronic structure with GGA, presented in the last section, predicts the AFI behavior even without U. To see whether the LDA really provides a metallic solution, we performed the calculations using a relatively highly dense k-mesh (~ 3350 irreducible k-points) and large number of plane waves (~ 22000) and we found a zero band gap as shown in Fig. 5a. A very small value of U opens up the gap. Therefore the metallic solution is to some extent a computational inaccuracy rather than a correlation problem.

To further investigate the effect of correlation, we have plotted U dependent DOS in Fig. 5b. As expected, with increasing U, the occupied d states are pushed below and unoccupied d states are pushed above to open up a larger band gap. Also with increasing U, the d states penetrate the O-p spectrum and induce a reverse hybridization with the latter. The band gap vs U plot of Fig. 5c shows that for small values of U, $E_g \sim U$ for LDA and $E_g \sim U + E'_g$ for GGA. Here $E'_g$ is the gap without U. Such a situation arises for extreme localization of the states. For large value of U (> 5 eV) $E_g$ saturates as O-p states occupy the valence band maximum. We note that experimental band gap is (~ 3.8 eV)[16] which is obtained theoretically when U is close to 4.5 eV.

It is important to compare the electronic structure of FeO and LFPO as in both the cases Fe is in 2+ charge states as well as in high spin state. The difference is in the structure. In FeO, the octahedral symmetry is present while it is completely broken in LFPO as discussed in the early part of this paper. In this context we have schematically summarized the electronic and magnetic structure of both the compounds in Fig. 6. The electronic structure of FeO is understood from the DFT results reported by Anisimov et al.[10] and Terakura et al.[43] Within LDA, the degenerate $t_{2g}$ and $e_g$ states of the $FeO_6$ octahedra have larger band width in FeO since the covalent p-d and d-d interactions are maximized in this case. $Fe^{2+}$ being in high spin state, the spin-down $t_{2g}$ band remains partially occupied. Therefore, without U, FeO will always have a metallic solution. On the contrary in LFPO, the five non-degenerate d states with negligible p-d and d-d covalent interactions resemble to the electronic states of clusters and are localized enough to produce a narrow gap at $E_F$. The Hubbard term in the Hamiltonian (Eq. 1) amplifies this gap. If the Hamiltonian involves both intra-orbital and inter-orbital onsite Coulomb repulsions, ordering of the d orbitals with respect to the Fermi level might also change along with the gap amplification.[41]

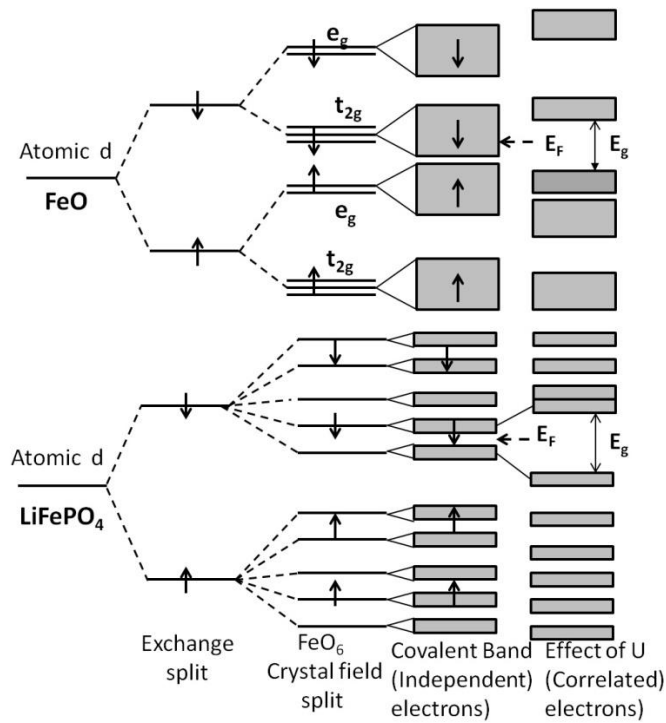

**Figure 6.** Schematic representation of the electronic structure of FeO (with perfect $FeO_6$ octahedra) and of LFPO (with distorted $FeO_6$ complex).

The mechanism that leads to the formation of AFI ground state in $LiFePO_4$ can be extended to other members of the $LiMPO_4$ (M = Cr, Mn, Co and Ni) (see Fig. 4 of the supplementary material of the published article).

## Magnetic Exchange Interactions and estimation of Neel temperature ($T_N$)

Since there are several possible antiferromagnetic configurations for LFPO and all of them exhibit insulating behavior, it is ncessary to study the stability of each magnetic ordering by calculating the spin exchange interactions (J). Experimentally, magnetocrystaline anisotropies are observed in the family of $LiMPO_4$ due to spin-orbit coupling.[44-46] For $LiFePO_4$ it is found to be moderate for ($\Delta g/g$ ~ 0.1, g is the Lande g factor).[44] This may be significant for the magnetoelectric effect[47] but the

Dzyaloshinskii-Moriya (DM) exchange interaction strength is found to be weak and its contribution to $T_N$ is very negligible for LFPO.[44,48] Therefore, we have neglected the DM term in this work. To estimate J in LFPO we have applied the extension of the Noodleman's broken-symmetry method[49] given by Dai and Whangbo.[50] According to this method, the energy difference between the high spin state and low spin state for a spin dimer can be approximated as:

$$E_{hs} - E_{ls} = \frac{1}{2}(S_{max})^2 J, \qquad (2)$$

where J is related to the spin-dimer Hamiltonian, $\hat{H} = J\,\hat{S}_1 \cdot \hat{S}_2$, with $S_{max}$ being the maximum spin of the dimer.

If both the sites of the dimer have same number of unpaired spins, say N, which is relevant for the LFPO, then Eq. 2 takes the form[50]

$$E_{hs} - E_{ls} = \frac{N^2}{4}J - \frac{-N^2}{4}J = \frac{N^2}{2}J. \qquad (3)$$

$E_{hs}$ and $E_{ls}$ are energies of the high and low spin states respectively which can be estimated from the DFT calculations as discussed below. In Eq. 3, N (= 4) represents the number of unpaired spins in each monomer. The net spin arrangement can be considered as a sum of individual spin-dimers of the lattice. The spin-dimers considered in this paper are shown in Fig. 7.

To evaluate the exchange interaction parameters from DFT results, we have considered seven magnetic configurations out of which one of them is ferromagnetic (FM). In other configurations (A1 – A6) at least one of the spin dimers is antiferromagnetic. The configurations for A1 (AFM(GS)) and A2 (AFM(ES)) are shown in Fig. 4 and are already discussed in the context of electronic structure. The rest are not shown in this paper, but can be mapped from the expression of their total spin- exchange energy which, using Eq. 3, can be written as:

$$\begin{aligned}
E_{FM} &= 4(2J_1 + J_2 + J_3 + J_4 + J_5 + J_6), \\
E_{A1} &= 4(-2J_1 + J_2 - J_3 + J_4 + J_5 + J_6), \\
E_{A2} &= 4(-2J_1 + J_2 + J_3 - J_4 - J_5 + J_6), \\
E_{A3} &= 4(-J_2 + J_4 - J_5 + J_6), \qquad (4) \\
E_{A4} &= 4(2J_1 + J_2 + J_6), \\
E_{A5} &= 4(J_2 + J_4 + J_5 - J_6), \\
E_{A6} &= 4(-J_2 - J_3 + J_6).
\end{aligned}$$

The energies, E, of Eq. 4 are now equated to the DFT calculated total energy of the respective magnetic configuration to estimate the $J_i$ values[51] and the results are listed in Table 3. Small value of $J_i$ (0 - 1 meV) suggest that LFPO is a magnetically weak system. Our estimated exchange interaction parameters are comparable with the previously reported values.[44,50] We attribute the weak magnetic interaction of the spin dimers to the extremely non-linear super exchange paths shown in Fig. 7. A pair of localized spins (dimer) always prefers to be antiferromagnetic which cannot be full filled when the spin-dimers are not isolated. For example an antiferromagnetic $J_1$ would prefer ferromagnetic $J_4$ and $J_6$ as can be observed from Fig. 7.

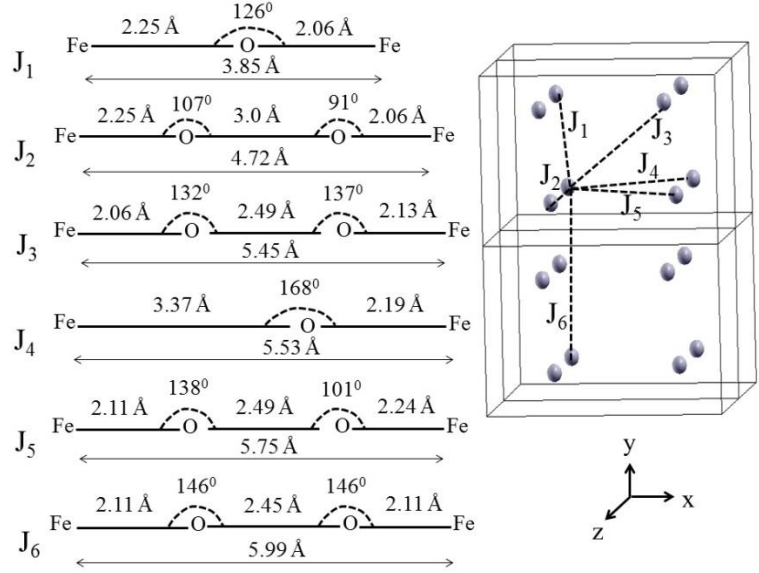

**Figure 7.** (Left) Various possible Fe - Fe spin dimers ($J_i$). For each them the super exchange path (Fe - O - Fe or Fe - O - O - Fe) is also shown. (Right) Mapping of $J_i$ in the crystal. For clarity only Fe atoms are presented.

Therefore, the system may exhibit incommensurate antiferromagnetic ordering. While so far it has not been observed experimentally in LFPO, related compounds $LiNi_{1-x}Fe_xPO_4$ have exhibited the incommensurate antiferromagnetic ordering.[52,53]

The Neel temperature ($T_N$) can be evaluated from the Curie-Weiss temperature[54] as follows:

$$\theta_{CW} = \frac{S(S+1)}{3K_B}\sum_i Z_i J_i \quad , \quad T_N = \frac{|\theta_{CW}|}{\mu} \qquad (5)$$

where $Z_i$ represents the number of equivalent magnetic neighbors corresponding to $J_i$ (see Table 3). $K_B$ is the Maxwell-Boltzmann constant and μ is a mean field constant. From the experimental studies[55] $\theta_{WC}$ and $T_N$ are found to be -115 K and 51 K respectively and hence μ ~ 2.25. Now taking S as 2, we have estimated $\theta_{WC}$ and $T_N$ for the optimized and experimental structures and the results are listed in Table 3.

Our DFT calculations predict the Neel temperature in the range 54.04 to 64.36 K which agree well with the experimental value. However, one has to be careful while predicting the exchange interactions from DFT. The mean-field based calculations, such as DFT, overestimate the exchange interaction strengths. Also we would like to note that in the present case, a change in the total energy by 1 meV can change the $T_N$ by 5 K on either side. For better estimation of exchange interactions and $T_N$, more comprehensive methods at the atomic scale such as DFT+DMFT[56] and atomistic spin dynamics[57] may be used.

| Structure | $J_1$ ($Z_1$) | $J_2$ ($Z_2$) | $J_3$ ($Z_3$) | $J_4$ ($Z_4$) | $J_5$ ($Z_5$) | $J_6$ ($Z_6$) | $\Theta_{CW}$ (K) | $T_N$ (K) |
|---|---|---|---|---|---|---|---|---|
| Optimized | 0.90 (4) | 0.04 (2) | 0.45 (2) | 0.02 (2) | 0.04 (2) | 0.27 (2) | -121.61 | 54.04 |
| Experimental | 0.96 | 0.03 | 0.64 | 0.06 | 0.04 | 0.43 | -144.82 | 64.36 |

**Table 3.** Exchange interaction parameters ($J_i$), in meV, are obtained by equating the DFT calculated energies to that of Eq. 4. The corresponding exchange paths are shown in Fig. 7. $Z_i$ represents the number of equivalent magnetic neighbors for $J_i$ exchange path. Positive and negative values of $J_i$ indicate antiferro and ferro ordering respectively. The Curie-Weiss temperature $\Theta_{CW}$ and the Neel temperature $T_N$ are evaluated using Eq. 5.

## SUMMARY

In summary, we have studied the electronic and magnetic structure of the cathode material $LiFePO_4$ with the objective to study the role of a polyanion in transition metal oxides. We find that the stable $(PO_4)^{3-}$ polyanion makes the compound non-planar and the degree of non-planarity determine the electronic and magnetic stability of the system. Here, the $FeO_6$ complex is formed to stabilize the $(PO_4)^{3-}$ polyanion. The former is largely distorted where the octahedral symmetry is completely absent. Therefore, the crystal field induced on the Fe-d orbitals is strongly asymmetric and splits the degenerate d states into multiple non-degenerate states. These states are localized, in the absence of strong d-d and p-d and p-p covalent interactions, and resemble to that of zero-dimensional systems such as clusters and nano-dots. The crystal field splitting is sufficient to introduce a narrow band gap which is unlikely for the conventional Mott insulators. The increase in Hubbard U simply enhances the magnitude of the gap. The spin-dimer analysis presented in this paper suggests a weak antiferromagnetic ordering and the estimated Neel temperature is found to be lying in the range [54 – 64K] which agrees very well with the experimental value of 51K. The mechanism presented in this article to explain the antiferromagnetic and insulating ground state of LFPO is extendable to other members of transition metal olivine phosphates and is expected to be a prototype for many transition metal oxides having polyanions.

## REFERENCES

1. Padhi, A. K., Nanjundaswamy, K. S., & Goodenough, J. B. Phospho-olivines as positive-electrode materials for rechargeable lithium batteries. *J. Electrochem. Soc.*, **144**, 1188-1194 (1997).
2. Goodenough, J. B. & Park, K. –S. The Li-ion rechargeable battery: a perspective. *J. Am. Chem. Soc.*, **135**, 1167-1176 (2013).
3. Nishimura, S. –I. et al. Experimental visualization of lithium diffusion in $Li_xFePO_4$. *Nat. Mater*, **7**, 707-711 (2008).
4. Islam, M. S. & Fisher, C. A. J. Lithium and sodium battery cathode materials: computational insights into voltage, diffusion and nanostructural properties. *Chem. Soc. Rev.*, **43**, 185-204 (2014).
5. Rousse, G., Rodriguez-Carvajal, J., Patoux, S. & Masquelier, C. Magnetic structures of the triphylite $LiFePO_4$ and of its delithiated form $FePO_4$. *Chem. Mater.*, **15**, 4082-4090 (2003).
6. Baker, P. J. et al. Probing magnetic order in $LiMPO_4$ (M = Ni, Co, Fe) and lithium diffusion in $Li_xFePO_4$. *Phys. Rev. B* **84**, 174403, 1-8 (2011).

7. Chung, S.-Y. , Bloking, J.T. & Chiang, Y.-M. Electronically conductive phospho-olivines as lithium storage electrodes. *Nat. Mater*. **1**, 123-128 (2002).
8. Delacourt, C. et al. Toward Understanding of Electrical Limitations (Electronic, Ionic) in $LiMPO_4$ (M = Fe, Mn) Electrode Materials. *J. Electrochem. Soc*., **152**, A913-A921, (2005).
9. Imada, M., Fujimori, A. & Tokura, Y. Metal-insulator transitions. *Rev. Mod. Phys*., **70**, 1039-1263 (1998).
10. Anisimov, V. I., Zaanen, J. & Andersen, O. K. Band theory and Mott insulators: Hubbard U instead of Stoner I. *Phys. Rev. B* **44**, 943-953 (1991).
11. Tang, P. & Holzwarth, N. A. W. Electronic structure of $FePO_4$ , $LiFePO_4$, and related materials. *Phys. Rev. B* **68**, 165107, 1-10 (2003).
12. Xu, Y. –N., Chung, S. –Y., Bloking, J. T., Chiang, Y. –M. & Ching, W. Y. Electronic structure and electrical conductivity of undoped $LiFePO_4$. *Electrochem. and Solid-State Lett*., **7** , (6), A131-A134 ( 2004 ).
13. Bacq, O. L. & Pasturel, A. First-principles study of $LiMPO_4$ compounds (M=Mn, Fe, Co, Ni) as electrode material for lithium batteries. *Philosophical Magazine*, **85**, 1747-1754 (2005).
14. Shi, S. et al. First-principles investigation of the structural, magnetic, and electronic properties of olivine $LiFePO_4$. *Phys. Rev. B* **71**, 144404, 1-6 (2005).
15. Zaghib, K., Mauger, A., Goodenough, J. B., Gendron, F. & Julien, C. M. Electronic, optical, and magnetic properties of $LiFePO_4$: small magnetic polaron effects. *Chem. Mater*.**, 19**, 3740-3747 (2007).
16. Zhou, F., Kang, K., Maxisch, T., Ceder, G. & Morgan, D. The electronic structure and band gap of $LiFePO_4$ and $LiMnPO4$. *Solid State Commun*, **132** , 181-186 (2004).
17. Kinyanjui, M. K. et al. Origin of valence and core excitations in $LiFePO_4$ and $FePO_4$. *J. Phys.: Condens. Matter* **22** , 275501, 1-8 (2010).
18. Mott, N. F. Metal – Insulator Transition. *Rev. Mod. Phys*., **40**, 677-683 (1968).
19. Zaanen J. & Sawatzky, G. A. Band gaps and electronic structure of transition-metal compounds. *Phys. Rev. Lett*. **55**, 418-421 (1985).
20. Slater, J. C. Magnetic effects and the HarLree-Fock equation. *Phys. Rev*. **82**, 538-541 (1951).
21. Reimann, S. M. & Manninen, M., Electronic structure of quantum dots. *Rev. Mod. Phys*., **74**, 1283-1341 (2002).
22. Ballhausen, C. J. Introduction to ligand field theory. (*McGraw-Hill*, New York, 1962).
23. Schuler, T. M. et al. Character of the insulating state in NiO: A mixture of charge-transfer and Mott-Hubbard character. *Phys. Rev. B* **71**, 115113, 1-7 (2005).
24. Arita, R., Kunes, J., Kozhevnikov, A.V., Eguiluz, A.G. & Imada, M. Ab initio Studies on the interplay between spin-orbit interaction and Coulomb correlation in $Sr_2IrO_4$ and $Ba_2IrO_4$. *Phys. Rev. Lett*. , **108**, 086403, 1-5 (2012).
25. Garcı´a-Moreno, O. et al. Influence of the Structure on the Electrochemical Performance of Lithium Transition Metal Phosphates as Cathodic Materials in Rechargeable Lithium Batteries: A New High-Pressure Form of $LiMPO_4$ (M = Fe and Ni). *Chem. Mater*. **13**, 1570-1576 (2001).
26. Vanderbilt, D. Soft self-consistent pseudopotentials in a generalized eigenvalue formalism. *Phys. Rev. B* **41** (Rapid Communications), 7892-7895 (1990).
27. Giannozi P. et al. QUANTUM ESPRESSO: a modular and open-source software project for quantum simulations of materials. *J. Phys., Condens. Matter* **21** , 395502, 1-19 (2009).
28. Perdew, J. P., Burke, K. & Ernzerhof, M. Generalized Gradient Approximation Made Simple. *Phys. Rev. Lett*. **77**, 3865-3868 (1996).
29. Shull, C. G., Strauser, W. A. & Wollan, E. O. Neutron diffraction by paramagnetic and antiferromagnetic substances. *Phys. Rev*. **83**, 333-345 (1951).
30. Roth, W. L. Magnetic structures of MnO, FeO, CoO, and NiO. *Phys. Rev*. **110**, 1333-1341 (1958).
31. Megaw, H. D. Crystal structure of Barium Titanate. *Nature*, **155**, 484-485 (1945).

32. Bhalla, A. S., Guo, R., Roy, R. The perovskite structure-a review of its role in ceramic science and technology. *Mat. Res. Innovat*. **4**, 3-26 (2000).
33. Siegrist, T., Zahurak, S. M., Murphy, D. W. & Roth, R. S. The parent structure of the layered high-temperature superconductors. *Nature*, **334**, 231-232 (1988).
34. Ruddlesden, S. N. & Popper, P. New compounds of the $K_2NIF_4$ type. *Acta Crystallogr*, **10**, 538-539 (1957).
35. Ruddlesden, S. N. & Popper, P. The compound $Sr_3Ti_2O_7$ and its structure. *Acta Crystallogr*, **11**, 54-55 (1958).
36. Satpathy, S., Popovic´ Z. S. & Vukajlovic, F. R. Electronic structure of the perovskite oxides: $La_{1-x}Ca_xMnO_3$. *Phys. Rev. Lett*. **76**, 960-963 (1996).
37. Nanda, B. R. K. & Satapathy, S. Electronic and magnetic structure of the $(LaMnO_3)_{2n}/(SrMnO_3)_n$ superlattices. *Phys. Rev. B* **79**, 054428, 1-6 (2009).
38. Goodenough, J. B. Coexistence of localized and itinerant d electrons. J. B. *Mat. Res. Bull*., **6**, 967-976 (1971).
39. Kimura, T. et al. Interplane tunneling magnetoresistance in a layered manganite crystal. *Science* **274**, 1698-1701 (1996).
40. Bednorz, J. G. & Muller, K. A. Possible High Tc Superconductivity in the Ba-La-Cu-O system. *Z. Phys. B* **64**, 189-193 (1986).
41. Craco, L. & Leoni, S. Electron localization in olivine materials for advanced lithium-ion batteries. *J. Appl. Phys*., **111**, 112602, 1-5 (2012).
42. Craco, L. & Leoni, S. Electrodynamics and quantum capacity of $Li_xFePO_4$ battery material. *Appl. Phys*. Lett., **99**, 19210, 1-3 (2011).
43. Terakura, K., Oguchi, T., Williams, A. R. & Kubler J. Band theory of insulating transition-metal monoxides: band-structure calculations. *Phys. Rev. B* **30**, 4734-4746 (1984).
44. Toft-Petersen, R. et al. Anomalous magnetic structure and spin dynamics in magnetoelectric $LiFePO_4$. *Phys. Rev. B* **92**, 024404, 1-9 (2015).
45. Tian, W., Li, J., Lynn, J. W., Zarestky, J. L. & Vaknin, D. Spin dynamics in the magnetoelectric effect compound $LiCoPO_4$. *Phys. Rev. B* **78**, 184429, 1-6 (2008).
46. Jensen, T. B. S. et al. Field-induced magnetic phases and electric polarization in $LiNiPO_4$. *Phys. Rev. B* **79**, 092412, 1-4 (2009).
47. Scaramucci, A., Bousquet, E., Fechner, M., Mostovoy, M., & Spaldin, N. A. Linear Magnetoelectric Effect by Orbital Magnetism. *Phys. Rev. Lett*. **109**, 197203, 1-5 (2012).
48. Liang, G. et al. Anisotropy in magnetic properties and electronic structure of single-crystal $LiFePO_4$. *Phys. Rev. B* **77**, 064414, 1-12 (2008).
49. Noodleman, L. Valence bond description of antiferromagnetic coupling in transition metal dimers. *J. Chem. Phys*. **74**, 5737-5743 (1981).
50. Dai, D. et al. Analysis of the spin exchange interactions and the ordered magnetic structures of Lithium transition metal phosphates $LiMPO_4$ (M = Mn, Fe, Co, Ni) with the olivine structure. *Inorg. Chem*., **44**, 2407-2413 (2005).
51. Oguchi, T., Terakura, K., & Williams, A. R. Band theory of the magnetic interaction in MnO, Mns, and NiO. *Phys. Rev. B*, **28**, 6443-6452 (1983).
52. Li, J. et al. Tweaking the spin-wave dispersion and suppressing the incommensurate phase in $LiNiPO_4$ by iron substitution. *Phys. Rev. B* **79**, 174435, 1-7 (2009).
53. Zimmermann, A. S., Sondermann, E., Li, J., Vaknin, D. & Fiebig, M. Antiferromagnetic order in $Li(Ni_{1-x}Fe_x)PO_4$ (x = 0.06, 0.20). *Phys. Rev. B* **88**, 014420, 1-7 (2013).
54. Kahn, O. Molecular Magnetism. Ch. 2, 26-29 (VCH: Weinheim, 1993).
55. Arcˇon, D., Zorko, A., Dominiko, R., & Jaglicˇ´ic´, Z. A comparative study of magnetic properties of $LiFePO_4$ and $LiMnPO_4$. *J. Phys.: Condens. Matter*, **16**, 5531-5548 (2004).
56. Kvashnin, Y. O. et al. Exchange parameters of strongly correlated materials: Extraction from spin-polarized density functional theory plus dynamical mean-field theory. *Phys. Rev. B*, **91**, 125133, 1-10 (2015).
57. Skubic, B., Hellsvik, J., Nordstrom, L., & Eriksson, O. A method for atomistic spin dynamics simulations: implementation and examples. *J. Phys.: Condens. Matter,* **20**, 315203, 1-12 (2008).

## ACKNOWLEDGMENT

This work was funded by NISSAN RESEARCH PROGRAM through grant no. PHY1314289NRSPBIRA. Institute HPCE facility was used for the computations. The authors would like to thank Sudakar Chandran for stimulating discussions.

**Author contributions statement**: B. R. K. N. conceived the idea. A. J. and B. R. K. N. performed the calculations and analyzed the results. The manuscript is prepared by B. R. K. N. and reviewed by A. J..

**Additional Informations:** The authors do not have any competing financial interests.